\RequirePackage{fix-cm}
\documentclass[twocolumn]{svjour3}          

\usepackage{epsfig}
\usepackage{amsfonts}
\usepackage{amsmath}
\usepackage{graphicx}
\usepackage{amssymb}
\usepackage{subfigure}
\usepackage{xcolor}
\usepackage{times}
\usepackage{hyperref}

\journalname{Cognitive Neurodynamics}

\begin{document}

\title{%
Desynchronizing effect of high-frequency stimulation in a generic
cortical network model
}


\author{Markus Sch\"utt
\and
Jens Christian Claussen
}


\institute{%
Markus Sch\"utt
\at
            University of Luebeck \\
            Institute for Neuro- and Bioinformatics\\
            D-23538 L\"ubeck \\
	    \and
	    Jens Christian Claussen (Corresponding author)
 \at
	    University of Luebeck \\
	    Institute for Neuro- and Bioinformatics\\
	    D-23538 L\"ubeck \\
	    \email{claussen@inb.uni-luebeck.de}\\
}

\date{Received: 6. July 2011 / Revised: 21.\ February 2012 / Accepted: 21.\
 March 2012}

\maketitle

\noindent
\fbox{
\href{http://www.springerlink.com/content/1871-4080/}{\bf Cognitive Neurodynamics, Volume 6, in print (2012)}
}
\\

\begin{abstract}
Transcranial Electrical Stimulation (TCES) and Deep Brain Stimulation (DBS) are two 
different applications of electrical current to the brain used in different areas of medicine. 
Both have a similar frequency dependence of their 
 efficiency,
with the most pronounced effects around $100$Hz. 
We apply superthreshold electrical stimulation, 
spe\-ci\-fi\-cally depolarizing DC current, 
interrupted at different frequencies,
to a simple
model of a population of
cortical 
neurons
which uses phenomenological descriptions of neurons by Izhikevich 
and synaptic connections on a similar level of sophistication. 
With this model, we are able to 
reproduce the optimal desynchronization around 100Hz,
as well as to predict 
the 
full
frequency dependence of the 
efficiency of desynchronization,
and thereby
to give a possible explanation for the action 
mechanism of TCES.
\end{abstract}



\section{Introduction}
Deep Brain Stimulation (DBS) has attained much attention during the past fifteen years as a modern treatment of miscellaneous neural and movement disorders, 
especially in the treatment of the 
symptoms
associated with Parkinson's disease (PD) \cite{Parkinson-Behandlung}, 
 namely 
tremor, rigidity and bradykinesia, but also for epilepsy \cite{Epilepsiebehandlung 1}, dystonia \cite{Dystonia-Behandlung 1,Dystonia-Behandlung 2,Dystonia-Behandlung 3}, 
and essential 
tre\-mor 
\cite{Tremorbehandlung}. 
DBS also has promising effects in the treatment of 
ob\-ses\-sive-\-com\-pul\-sive 
disorder (OCD) \cite{OCD-Behandlung}, 
Tourette's syndrom \cite{Tourette-Syndrom 1,Tourette-Syndrom 2} and depression \cite{Depressionsbehandlung}. 
For the purpose of the treatment, 
electrodes are chronically implanted in a specific area of the brain. In the majority of cases
 of movement disorders, 
the targeted region is the subthalamic nucleus (STN), as the stalling of dopaminergic (inhibitory) neurons in the substantia nigra 
leads to pathological synchronized oscillations in the STN, which are correlated with the clinical symptoms of PD 
\cite{Absterben dopaminerger Neuronen}. 
Also other regions of the brain have been targeted for DBS,
the globus pallidus internus (GPi) 
for the treatment of PD \cite{DBS im GPi,functional block} 
 as well as
treatment of dystonia \cite{Dystonia-Behandlung 2,Dystonia-Behandlung 3},
 was attempted 
in the CA1 region of the hippocampus \cite{axonal conduction,Hippocampus-Epilepsie}, 
and in the ventral intermediate nucleus of the thalamus (Vim) \cite{functional block}. 
Frequencies between $100$ and $200$ Hz 
(clinically often $130$~Hz are applied)  
have been proven to give the best results in alleviating the symptoms, 
whereas low frequencies of roughly $10$Hz can even worsen the symptoms. 
Understanding the working mechanisms of DBS from computational models
is not an easy exercise:
Recurrent neural networks with inhibitory neurons can exhibit
rich behavior including synchronization
\cite{liu2010,zhang2010,wang2011};
various mechanisms of desynchronization by stimulation 
\cite{hauptmanntass2010}
or nonlinear feedback 
\cite{schollPTRSA2009}, 
to mention a few,
have been proposed.
Although there has been a lot of research, 
the action mechanism of DBS still remains elusive.

At least in the last years, however, 
less 
effort has been made to analyze the action mechanisms of transcranial electrical stimulation, 
which is used for different purposes and is labelled with different names. 
Probably the most established one is Transcranial Electrical Stimulation (TCES), 
a technique which has been used to 
reduce drug requirements for anaesthesia in surgical operations \cite{review 1998}. 
Although it has been used in over $10000$ operations at least up to 1998 \cite{review 1998}, 
its 
 working 
mechanism 
still remains to be explained.

The TCES technique was 
developed 
by Limoge in the 1970s \cite{Limoge}
and the technical protocol as well as electrode placement 
is established under the terminus {\sl Limoge current} \cite{review 1998}.
Although the way of the Limoge current through the head is still 
not precisely known,
and it is even not known 
if the current may, at least in parts, act by influencing peripheral nerves outside the 
cranium \cite{Zaghi der nirgendwo publiziert hat}, 
the part of the brain where transcranial electrical stimulation has its greatest effect is
presumably 
the neocortex,
because current density decreases with the distance to the electrode. 
Since the neocortex is the part
 of the brain 
closest to the stimulation for 
 all standard  
electrode positions, 
it is
quite likely the part carrying the greatest
 fraction
of the current.
Transcranial direct current
stimulation in humans
was also shown to enhance excitability
\cite{antal2004ophtal,antal2004eurjneurosci}
which on the one hand indicates plasticity effects
and on the other hand shows that TCES can have a 
pronounced effect on neocortical regions
and not only on specific subcortical structures
as targeted by DBS.

To account for an explanation of the action mechanism of TCES, we have built up a 
model of a population of neurons from
the mammalian cortex, and applied external stimulation with different frequencies. 
This model is presented in section \ref{Modellpraesentation}. 
In section \ref{verallgemeinertes Phasen-Konzept} we give a 
short introduction to generalized phase definitions and 
introduce order parameters which can be used to quantify 
the phase synchronization of the neurons. 
Results are presented in section \ref{Resultate} 
and discussed in section \ref{Diskussion}. 
In the final section \ref{Zusammenfassung} 
we give a comprehension and present possible extensions of the model.

\section{The model\label{Modellpraesentation}}
As we want to demonstrate the basic effects in a model that is as simple as
possible, 
we do not attempt to account for biological parameter variability, as we
expect that the basic mechanism does not rely on
detailed properties of a specific cortical region.
\subsection{Neuron model}
Our model cortex consists of $N_{exc}=1024$ excitatory and $N_{inh}=256$ inhibitory neurons. 
This reflects the fact that the ratio of excitatory to inhibitory neurons 
(in the 
mammalian cortex)
is $4:1$ 
\cite{PING,izhi_Gehirn}. Individual neurons were modelled according to the 
Izhikevich model of spiking neurons \cite{izhi}, which contains two variables $v$ and $u$, 
representing 
the membrane potential, and a recovery variable, 
e.g.\ a slow K$^{+}$ current, respectively. 
$v$ and $u$ are governed by the differential equations
\begin{eqnarray}
\dot v&=&0.04v^2+5v+140-u+I_{ext}+I_{syn}
\mbox{~~~~}
\\
\dot u&=&a(bv-u),
\end{eqnarray}
where $I_{ext}$ and $I_{syn}$ represent the external and the synaptic input current, respectively. 
If $v\geq30$, 
a reset is initiated,
\begin{equation}
v\to c{\rm ,}\hphantom{++}u\to u+d 
\end{equation}
Depending on the four parameters $a$, $b$, $c$, and $d$, the Izhikevich neuron can map a rich variety of neuronal spiking patterns. We use $a=0.02$, $b=0.2$, $c=-65$, $d=8$ for the excitatory subpopulation and $a=0.1$, $b=0.2$, $c=-65$, $d=2$ for the inhibitory subpopulation. These parameter values correspond to 'regular spiking' (RS) and the 'fast spiking' (FS) pattern, which are exhibited by most of the excitatory and of the inhibitory neurons in the cortex, respectively 
\cite{pyramidal sind RS,izhi_Gehirn}.

\subsection{Modeling synapses}
Each neuron makes a synaptic connection to any other neuron with probability $p=200/(N_{exc}+N_{inh})$, 
so each neuron has $200$ synaptic connections on average. 
In reality, there are thousands of synapses per neuron \cite{izhi_Gehirn},
 whereof here we model a small subnetwork of a local cortical assembly and 
 therefore cannot explicitely consider long-range connections.
 Consequently, we also neglect structured connectivity, apart 
 from taking into account a sparse random connectivity.
The synaptic input current to a neuron $i$ is given by the sum over the postsynaptic potentials of all neurons $j$ presynaptic to $i$
\begin{equation}
I_{syn,i}=\sum_{j=1}^{N_{exc}+N_{inh}} \; L_{j,i} \; s_{x_j}\; g_{x_j,x_i}(t_j-\delta) 
\;
e^{-\frac{t_j-\delta}{\tau_{x_j}}} 
\end{equation}
where $x_j$ and $x_i$ are the types (i.e. excitatory or inhibitory) of the pre- and the 
postsynaptic neurons, respectively, and can take on the values $E$ and $I$. 
Here 
$(L_{j,i})$ is the adjacency matrix describing whether synaptic connections between $j$ and $i$ exist. 
The sign variables $s_{x_j}$ formally account for the excitatory or inhibitory nature of the presynaptic neuron, with $s_E=+1$ and $s_I=-1$. 
Then, the $g_{x_j,x_i}$ are the synaptic strengths between two types of neurons, with $g_{E,E}=0.6$, $g_{E,I}=0.1$, $g_{I,E}=0.2$ and $g_{I,I}=0.05$. 
We have adapted the ratios of these values
according to the reference model
  \cite{Compte}. 
For the axonal delay we use $\delta=0.25$ms;
 $\tau_{x_j}$ is a time constant depending on the presynaptic neuron, with
$0.2$ms for excitatory and $0.4$ms for inhibitory neurons,
respectively,
and 
 $t_j$
denotes
 the time since the last spike of the presynaptic neuron $j$ (in ms).

To consider the dynamics of a local neural assembly 
under stimulation, it is
 of low relevance
from where incoming projections originate.
To ensure non-trivial activity in the network, external input to the neural network
 is assumed from 
 $N_{ext}=128$ external excitatory neurons. 
These neurons are not explicitly modelled, but they make synaptic connections into the network with the same properties as the 
network synapses, and fire action potentials at random with probability $0.01$ during each time step.
The model was integrated using a
fourth-order explicit Runge-Kutta method with a time 
step
 of $0.05$ms.

\section{Phase synchronization\label{verallgemeinertes Phasen-Konzept}}
 To quantify synchronization of an ensemble of oscillators, more convenient
 observables than correlation functions can be used.
 Here we follow an established approach \cite{Kuramoto,original_Kuramoto}
 to define a complex-valued order parameter based on a generalized 
 phase in a phase space appropriately chosen for the model at hand.
\subsection{Generalized phase}
There are at least three concepts for a generalized phase definition 
of chaotic oscillations \cite{phase definition}.
For the sake of numerical simplicity, 
we use a geometrical phase angle definition.
In absence of external stimulation, each neuron has an attracting
limit cycle in the $\{\vec{v},\vec{u},\vec{\dot v},\vec{\dot u}\}$-space;
to define a phase, this trajectory has to be suitably projected into a two-dimensional 
plane, which indeed is possible here.
We define the phase of the oscillator $i$ (i.e., the neuron $i$) as the angle between a 
given 
 (fixed)
direction and the 
 position of the neuron's state
in a chosen plane of the phase space,
\begin{equation}
\phi_i(t)=\arctan\frac{-(\dot v_i(t)-\dot v_{i,c})}{v_i(t)-v_{i,c}(t)},
\end{equation}
where the point $(v_{i,c},-\dot v_{i,c})$ is within the rotation centre of the attractor
of the neuron $i$ in the plane of the phase space spanned by $v_i$ and $-\dot v_i$, 
with $\dot v_i$ being the time derivative of the membrane potential $v_i$ of the neuron $i$.

The first choice for such a plane in the phase space to be used for phase definition 
would be the $\{\vec{v},\vec{u}\}$-space, but unfortunately the attractor in this plane 
has multiple rotation centres and changes its position for different stimulation intensities, 
so the $\{\vec{v},-\vec{\dot v}\}$-space is a better choice. 
To ensure a monotoneous increase of $\phi$, one has to choose a coordinate
centre within the trajectory loop;
here we select the point $(v_{i,c},-\dot v_{i,c})=(c_i,0)$
as rotation centre.
We further choose  $\{\vec{v},-\dot{\vec{v}}\}$ 
(rather than $\{\vec{v},\dot{\vec{v}}\}$ which results in $\dot{\phi}<0$) 
to ensure $\dot{\phi}>0$, for convenience.
 Other linear combinations of $\vec{v}$, $\vec{u}$ and their first time
 derivatives could be used as well as an embedding of the dynamics.

\subsection{Order parameter}
We define the complex-valued order parameter $\tilde r$ of the phases as
\begin{equation}
\tilde r:=re^{i\psi}=\frac{1}{N}\sum_{i=1}^{N}e^{i\phi_i}.
\end{equation}
This definition is according to the definition of the order parameter in the Kuramoto model of phase synchronization 
\cite{Kuramoto,original_Kuramoto}. We analyze the synchronization in the whole population 
($\tilde r_W$), the excitatory subpopulation ($\tilde r_E$) and the inhibitory subpopulation ($\tilde r_I$) as well, so $N$ can be $N_{exc}+N_{inh}$, $N_{exc}$ and $N_{inh}$, respectively. The absolute value $r$ of $\tilde r$ lies in the interval $[0,1]$, where $0$ corresponds to a complete unsynchronized state and $1$ to complete synchronization.

As these order parameters oscillate already for the unstimulated system, we are mainly interested in the average of the order parameters, which we denote by a bar, e.g. 
$\overline{r}_W$.

\section{Results\label{Resultate}}
\subsection{Behaviour of the unstimulated system}
In the unstimulated system, we observe tightly synchronized spike volleys that occur with $\sim 10$Hz. 
This is in the alpha-range which is associated with an alert but relaxed state. 
\cite{readingEEGs}.
This synchronization phenomenon is 
widely 
known for random coupled neural networks 
\cite{grosser raddrehender koenig,izhi,Polychronization,PING}.

As one can already suspect by looking at the distribution of the spikes over time, all three order parameters have big values close to $1$, at least between spike volleys. When such a population spike occurs, the order parameters break down and reach values that can be as small as $0.2$ or even smaller for a very short time.

\begin{figure}
\begin{center}
    \includegraphics[width=140mm]{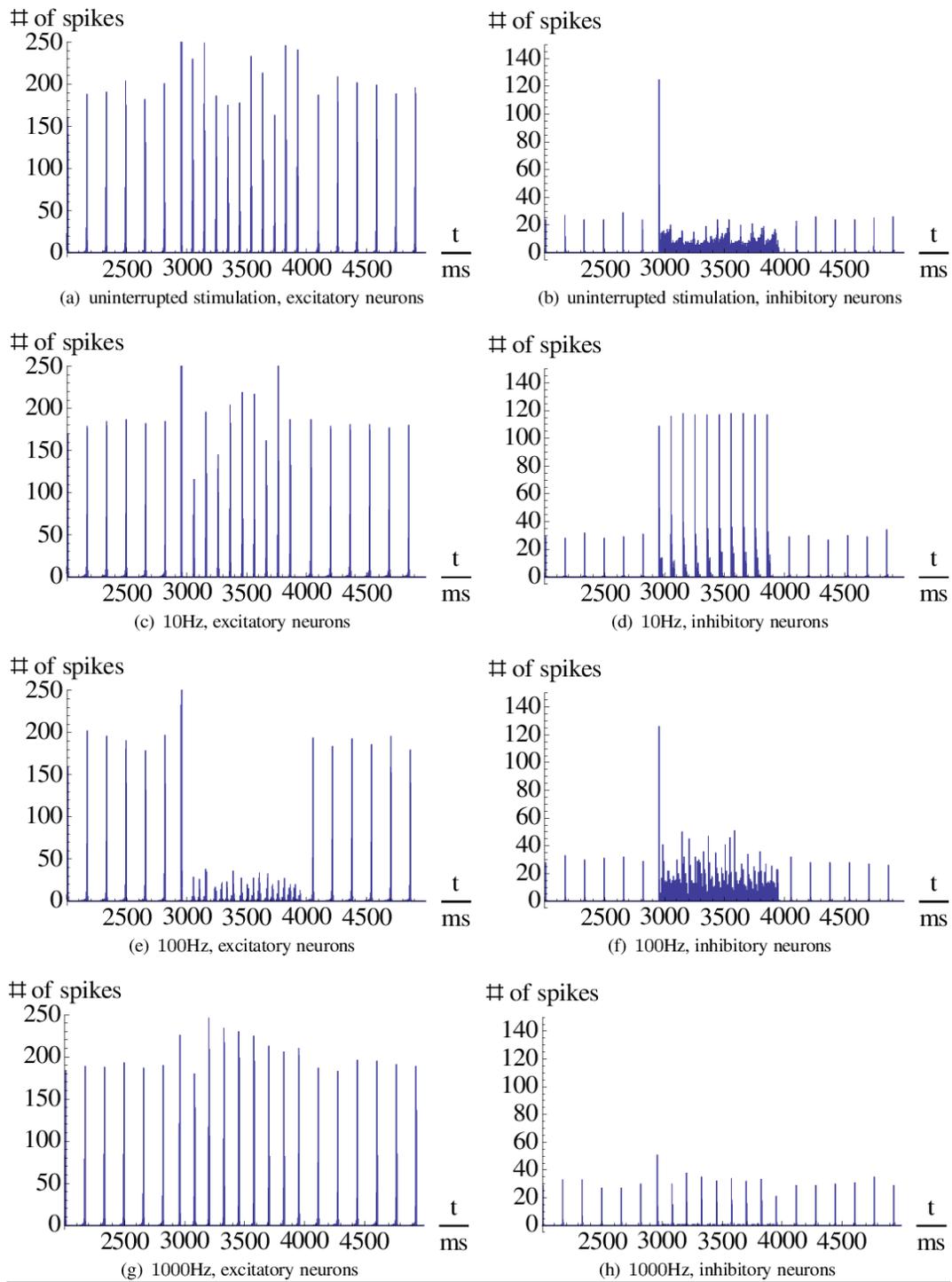} 
\end{center}
\caption{Spike histograms of $3000$ms of four simulations, 
for the excitatory (left-hand side) and the inhibitory (right-hand side) subpopulation. 
The stimulation protocol is as follows:
Stimulation $I_{ext}$ was switched on $3000$ms after initialization for
$1000$ms.
~~Each stimulation is monopolar (excitatory DC stimulation) and
either uninterupted DC (a,b), or modulated by a rectangular envelope of
10Hz (c,d), 100Hz (e,f), or 1000Hz (g,h), respectively.
Before and after the stimulation, the behavior of the unstimulated system
is visible.
\label{fig1spikehisto}
}
\end{figure}

\begin{figure}
\begin{center}
\includegraphics[width=120mm]{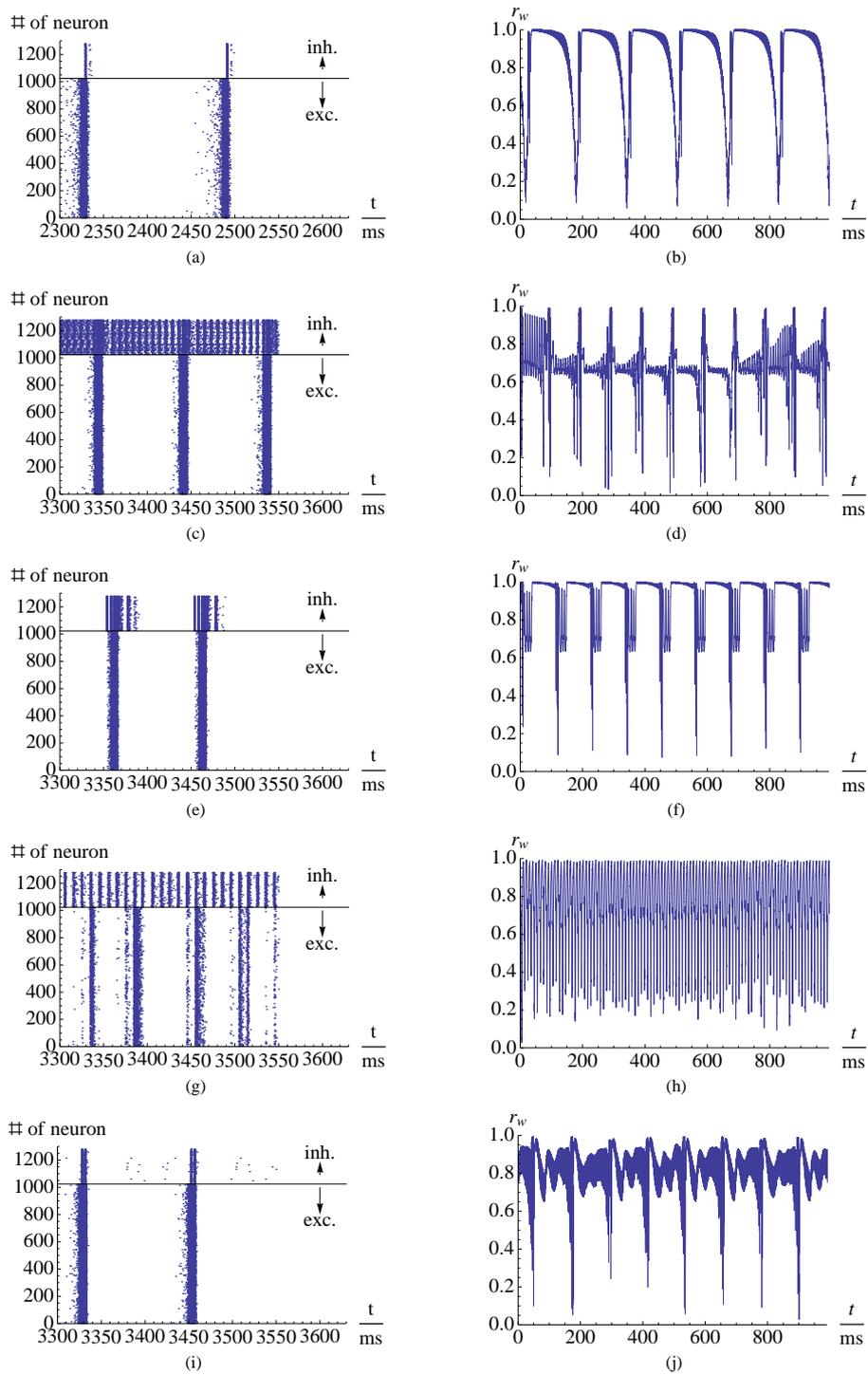}
\end{center}
\caption{Left-hand side: Spike times of the neurons in the network. Each point denotes a spike. 
The ordinate denotes the number of the neuron if numbered serially. Above the black line are the inhibitory neurons, below it the excitatory neurons. 
Right-hand side: The order parameter 
$r_W$ over the time (ordinata: relative to stimulation onset $t=0$). 
Amplitude of stimulation is $10$ in both cases. Note the different scaling of the abscissae on the left- and the right-hand side. (a),(b) before stimulation. (c),(d) uninterrupted stimulation. (e),(f) $10$Hz rectangular pulses. (g),(h) $100$Hz rectangular pulses. (i),(j) $1000$Hz rectangular pulses.
\label{fig2}
}
\end{figure}

The dynamics of the unstimulated system is shown in Fig.~\ref{fig1spikehisto} 
in each of the subfigures (a-h) before the stimulus onset.
At 3000ms,  stimulation impulses were applied according to
different protocols as described below.
For each stimulation type, Fig.~\ref{fig2} 
shows the respective network activities (left panels), as well as the time-dependence of
the order parameter (right panels).

\subsection{Uninterrupted DC stimulation}
For (uninterrupted) DC stimulation, the frequency of the spike volleys is increased, but the spike volleys of the excitatory subpopulation still remain discriminable. In contrast to that, the inhibitory subpopulation now fires continuously, as there are only few and relatively weak inhibitory-inhibitory synapses which could affect this subpopulation to 
cease fire.

\onecolumn

\twocolumn

For this type of stimulation, 
$\overline{r}_W$ is decreased. In fact, it is the strongest reduction of
$\overline{r}_W$ that we observe for all stimulation frequencies. But if one takes a view on the order parameters of the two subpopulations, one gets a different finding. $\overline{r_E}$ is even increased. In contrast, $r_I$ fluctuates strongly and fast almost between $1$ and $0$. Therefore, $\overline{r_I}$ is decreased all in all, but not as strong as 
$\overline{r}_W$. So, the reduction of 
$\overline{r}_W$ stems in parts from the desynchronization of the inhibitory subpopulation, but also from the 
desynchronization of both subpopulations 
from each other.
\subsection{Low frequency stimulation ($10$Hz)}
\begin{figure}
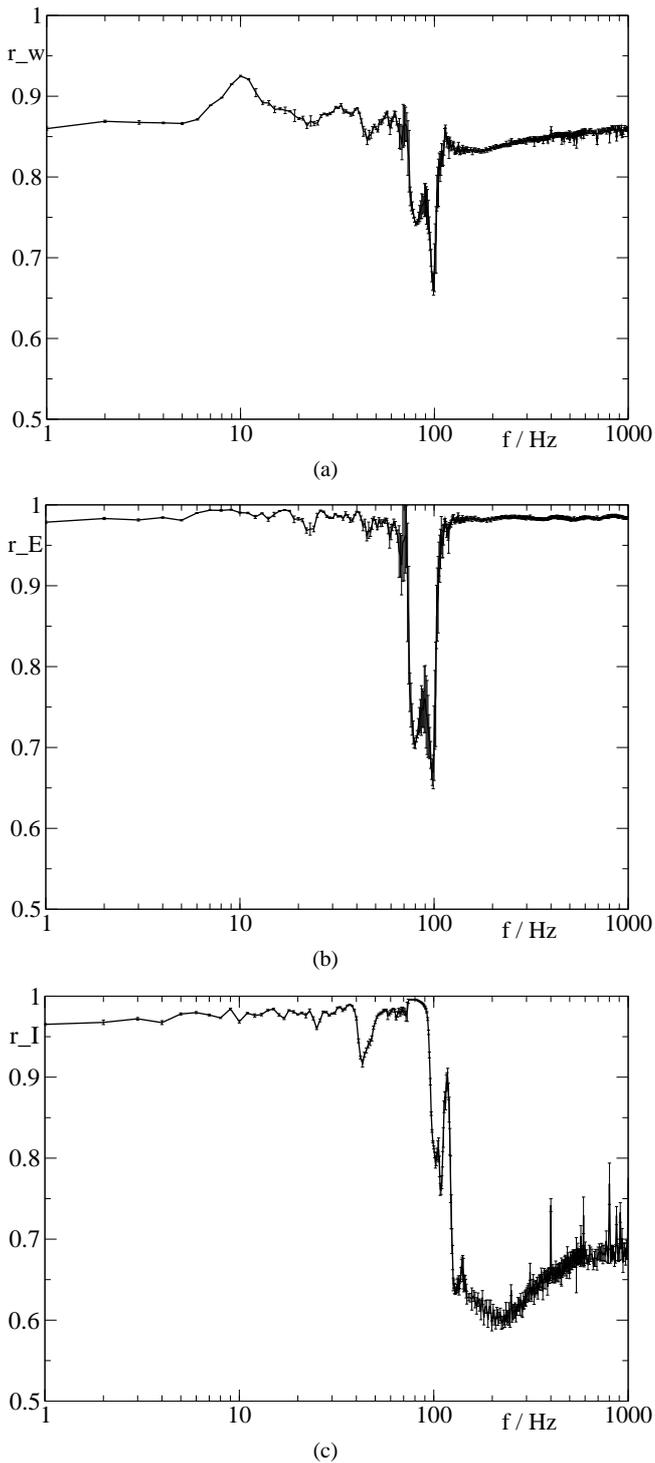

\centering
\subfigure[]{\includegraphics[width=85mm]{run1000allm1.eps}}
\subfigure[]{\includegraphics[width=85mm]{run1000allm2.eps}}
\subfigure[]{\includegraphics[width=85mm]{run1000allm3.eps}}
%
\caption{Dependence of the order parameter on the frequency of stimulation. Intensity of stimulation is $10$. 
(a) $r_W$ of the whole population. (b) $r_E$ of the excitatory subpopulation. (c) $r_I$ of the inhibitory subpopulation. 
In all three plots, averages are shown over 15 simulations, and the
corresponding standard deviations are indicated by error bars.
\label{fig3}
}
\end{figure}

For low frequency stimulation, the spiking pattern of the excitatory subpopulation remains almost unchanged, but as such frequencies are near the systems eigenfrequency, 
we have a resonance phenomenon. 
Therefore, the population spikes are now time-locked to the stimuli, and occur with exactly the freqency of the stimulation. 
 Hence the activity becomes more synchronized,
 leading to the observable effect that
all three order parameters are increased
 (Fig.\ \ref{fig3}).

\subsection{High frequency stimulation ($100$Hz)}
If we increase the frequency of the stimulation to high frequencies of $\sim 100$Hz, 
the spike volleys still occur during stimulation and
synchronize with the stimulation frequency. 
If the amplitude of the stimulation is not high enough, 
some stimuli may be missed, so that, for example, 
three spikes occur locked to the stimuli, and then for another two stimuli, the neural network is silent.

This finding is supported by the order parameter $\overline{r}_W$, which decreases around $100$Hz. It does not decrease as much as for the uninterrupted stimulation, but in contrast to that kind of stimulation, the order parameter
$\overline{r}_E$ of the excitatory subpopulation decreases as well, and below the baseline value of the unstimulated case. Similarly,
$\overline{r}_I$ is decreased, but the minimum is not as sharp as for
$\overline{r}_E$.

\subsection{Very high frequency stimulation ($1000$Hz)}
If we increase the stimulation frequency further, the individual neurons cannot follow the stimulation frequency, and we observe spike time characteristics that are very similar to the unstimulated case. All three order parameters again increase to higher values. One could think that it is possible to reproduce the 
desynchronizing effect of the moderate high frequencies by increasing the stimulus amplitude, 
but unfortunately
we could not generate such effect (not shown).

\subsection{Post-stimulation characteristics}
As we did not model aspects like synaptic plasticity, there are no long-term effects of the stimulation. All effects cancel out with cessation of stimulation, except a small reduction in spike rate, which soon returns to baseline values.

\section{Discussion\label{Diskussion}}
We applied depolarizing electrical stimulation to a simple cortex model. The stimulating current is interrupted with different frequencies.

If not stimulated, the model exhibits synchronized spike volleys or population spikes at frequencies that lie in the 
alpha range. This synchronization phenomenon is 
widely
known
for randomly coupled neural networks \cite{grosser raddrehender koenig,PING,izhi,Polychronization}. The three order parameters we have defined are close to the maximum value $1$ 
when there are no spike volleys, but break down when population spikes occur. 
This is not surprising as the movement on the trajectories is very fast during spikes, 
compared to the time between spikes.
Thus, if there is only a little difference between the spike timings of two different neurons, they may be far apart from each other, whereas for the time 
between population spikes, all neurons have values of $v$ and $-\dot v$ located in a small volume of the phase space. Thus the phases are very similar between population spikes and more different during the population spikes, 
and the order parameters break down in the latter case.

For uninterrupted DC stimulation the frequency of the spike volleys increases, and the inhibitory subpopulation begins to fire continously. Although the average order parameter 
$\overline{r}_W$ of the whole population decreases strongly, the order parameter
$\overline{r}_E$ of the excitatory subpopulation is increased by uninterrupted stimulation. Enhanced synchronization in large groups of neurons 
is also reason for epileptic spasms. 
As motor neurons (which are - excitatory - pyramidal neurons) 
can be effected by TCES as well as any other population in the cortex, we think this effect
could be a possible explanation of the clonic spasms which are 
produced by uninterrupted high intensity extracranial DC stimulation \cite{Limoge}.

If the stimulation is interrupted at low frequencies, population spikes become time-locked to the stimuli, and the order parameters
$\overline{r}_W$ and $\overline{r}_E$ even increase. This is not unexpected as stimulation near the eigenfrequency of a system leads to resonance. 
If we think about DBS, 
for which 
 a similar frequency dependence of effectiveness 
as for TCES is observed
(e.g., compare \cite{Limoge,review 1998,Sances} for TCES with
\cite{Absterben dopaminerger Neuronen,uncovering mechanisms} 
for DBS), 
it is known that stimulation with low frequencies can even worsen the symptoms which are associated with too strong pathological synchronization \cite{worsen symptoms}. In terms of TCES, such a stimulation would possibly counteract any anaesthetic action, but this cannot be said definitely as there seem to be no experiments on that yet.

For stimulation with moderately high frequencies, we found the introduction of a new activity pattern. As the neurons try to follow the stimulation frequency, smaller spike volleys occur which are locked to the stimuli. But due to the short duration of the stimuli, some of them are missed by the population spikes, and only a smaller number of neurons participates in each such spike volley, compared to the intrinsic volleys. Effectively, this is a desynchronizaton of the neural activity. It is well known that Deep Brain Stimulation of the subthalamic nucleus with such moderately high frequencies cancels out the pathological synchronization induced by 
Parkinson's disease. 
We propose that in the cortex this desynchronization phenomenon intercepts neural information processing, as it is known from experiments on electroanaesthesia (EA) \cite{Limoge}, which eventually lead to the invention of TCES.

For very high stimulation frequencies, the desynchronizing effect again vanishes. This is in compliance with investigations in DBS \cite{axonal conduction}, reporting stimulation in 
the kHz range to have no effect, and with old reports on EA showing the current intensity necessary for the loss of consciousness to greatly increase for such high frequencies \cite{Sances}. In fact, increasing the stimulation amplitude could not fully compensate the effect of increasing the frequency in our model. Instead, by increasing the amplitude of stimulation, we rather found effects similar to those of uninterrupted current. Perhaps this explains why 
currents with a frequency of approximately $100$Hz, 
modulated with ultra high frequency currents ($\sim 100$kHz), 
are
comparatively 
effective than the
 unmodulated current
 \cite{Limoge,review 1998}. 
If the ultra high frequency current acts similar to an uninterrupted current, 
the effect is expected to be similar to the unmodulated $100$Hz current, 
but it has lesser secondary effects, as less current is delivered to the brain.
\\[-8mm]

\section{Summary and Outlook\label{Zusammenfassung}}
Deep Brain Stimulation and Transcranial Electrical Stimulation are two promising applications of 
electrical currents in medical practice. Whereas DBS has largely replaced pallidotomy in the 
treatment of Parkinson's disease 
\cite{uncovering mechanisms}, and is prosperous in the treatment of some other neural diseases \cite{Tremorbehandlung,Dystonia-Behandlung 1,OCD-Behandlung,Epilepsiebehandlung 1,Dystonia-Behandlung 2}, 
TCES yet does not have
 reached a comparative level of potential applications that
 it could have in surgical practice.

In summary, 
we have presented a generic cortex model which reproduces the frequency dependence of the activity of
TCES:
With uninterrupted stimulation, 
the overall synchronization is decreased but the synchronization 
of the excitatory subpopulation
is increased.
For moderately high frequencies synchronization of the whole 
population and the excitatory subpopulation is decreased as well. 
We assume the same mechanism to account for the similar 
frequency dependence of the efficacy of DBS, resulting in a 
drastic desynchronization of the spiking activity. 
Stimulation of our cortex models with unipolar $100$Hz rectangular waveforms leads to distinct changes of the spiking properties of the neural network, and it markedly reduces the  synchronization of the network, which is quantified by the order parameters
$\overline{r}_W$, $\overline{r}_E$ and $\overline{r}_I$. Whereas slower or higher stimulation frequencies do not lead to such distinct changes, 
and
instead can even tend to enforce the natural behaviour of the model, 
moderately high frequencies 
(around 100 Hz) 
lead to a desynchronization of activity. We therefore suppose that desynchronization of cortical activity and the introduction of cortical noise is at least in parts accountable for the effects of TCES, as it disrupts ongoing signal processing in the cortex.

This model obviously just gives hints to the the question how TCES actually works. 
To consolidate our theory that the desynchronization of cortical activity leads to the
beneficial effects of TCES we would have to build up a more complex and of course larger model containing more neurons, which should take into account the 
spatial structure of the brain. 
If a spatial distribution of the neurons is modeled, one should desist from assuming homogenous electrical fields and model explicitly the electric field distribution.
Also, neurons could be modeled in different degree of detail, e.g., 
according to Izhikevich \cite{izhi} taking into account more firing patterns, or as advanced versions of Hodgkin-Huxley neurons with realistic synaptic dynamics, including synaptic plasticity and axonal delays that depend on the spatial distances of the neurons.
Similarily, 
a multi-compartment model -- 
containing
at least two or three compartments for axon,
soma, and eventually dendrite -- could be used.
While the quantitative dynamics of such more detailed models might come
closer to reality,
we expect those models to exhibit a similar desynchronization effect in the
100Hz regime, as demonstrated in our model.
\\[2mm]
{\sl Acknowledgments:} 
Financial support by the Deutsche For\-schungs\-ge\-mein\-schaft
(DFG SFB-654 project A8 and
Graduate School for Computing in Medicine and Life Science)
is gratefully acknowledged.
\\[-8mm]

\end{document}